\documentclass[sigconf]{acmart}
\renewcommand\footnotetextcopyrightpermission[1]{}
\AtBeginDocument{%
  }

\acmConference[ASPDAC '27]{Make sure to enter the correct
  conference title from your rights confirmation email}{Jan 25--28,
  2027}{Tokyo}

\begin{document}
\title{UnO-NeSy: Optimizing Neuro-Symbolic Computing with Unified FeFET Based In-Memory System\\
}



\author{Yifei Zhou}
\email{jimzhou_28@zju.edu.cn}
\affiliation{%
  \institution{Zhejiang University}
  \city{Hangzhou}
  \country{China}
}

\author{Zhiying Zhu}
\email{	zy.zhu@zju.edu.cn}
\affiliation{%
  \institution{Zhejiang University}
  \city{Hangzhou}
  \country{China}
}

\author{Yuxiao Jiang}
\email{jiangyx@zju.edu.cn}
\affiliation{%
  \institution{Zhejiang University}
  \city{Hangzhou}
  \country{China}
}

\author{Yihan Mao}
\email{ianmao@zju.edu.cn}
\affiliation{%
  \institution{Zhejiang University}
  \city{Hangzhou}
  \country{China}
}

\author{Jiaqi Zhang}
\email{3220104397@zju.edu.cn}
\affiliation{%
  \institution{Zhejiang University}
  \city{Hangzhou}
  \country{China}
}

\author{Zhiguo Shi}
\email{shizg@zju.edu.cn}
\affiliation{%
  \institution{Zhejiang University}
  \city{Hangzhou}
  \country{China}
}

\author{Yuxiao Yang}
\email{yuxiao.yang@zju.edu.cn}
\affiliation{%
  \institution{Zhejiang University}
  \city{Hangzhou}
  \country{China}
}

\author{Zheyu Yan}
\authornote{Corresponding authors.}
\email{zyan2@zju.edu.cn}
\affiliation{%
  \institution{Zhejiang University}
  \city{Hangzhou}
  \country{China}
}

\author{Cheng Zhuo}
\authornotemark[1]
\email{czhuo@zju.edu.cn}
\affiliation{%
  \institution{Zhejiang University}
  \city{Hangzhou}
  \country{China}
}

\author{Xunzhao Yin}
\authornotemark[1]
\email{xzyin1@zju.edu.cn}
\affiliation{%
  \institution{Zhejiang University}
  \city{Hangzhou}
  \country{China}
}

\begin{abstract}
Neuro-symbolic artificial intelligence (NeSy AI) combines neural perception with symbolic reasoning, enabling logical reasoning.
Hyperdimensional computing (HDC) is promising for NeSy AI, owing to its high parallelism and strong robustness to noise.
However, the dense neural operations and memory-intensive symbolic inference in NeSy impose a severe "memory wall" bottleneck.
Existing NeSy accelerators either focus solely on symbolic kernels or rely on conventional digital architectures, lacking a unified and energy-efficient hardware substrate.
Moreover, exploring optimal hardware design configurations often requires costly trial-and-error.
In this work, we propose \textbf{UnO-NeSy}, a unified CiM framework for optimizing NeSy computing. 
UnO-NeSy consists of \textbf{UnO-CiM}, an ultra-low-power NeSy architecture based on a homogeneous
2FeFET--2R compute-in-memory (CiM) array that supports both neural
computation and HDC reasoning, and \textbf{UnO-Sim}, a cross-workload, end-to-end CiM simulator that models device non-idealities, neural quantization effects, and HDC execution behavior for accurate system-level evaluation.
Built on UnO-Sim, we further develop an automated optimizer to identify workload-specific operating points under coupled energy, accuracy, and hardware constraints.
By optimizing the operating configuration of the baseline 2FeFET--2R
design, UnO-NeSy achieves $61.8\%$ higher energy efficiency and $38.2\%$
lower full-system inference energy.
Across seven RAVEN configurations, UnO-NeSy achieves $96.2\%$ average
reasoning accuracy under hardware non-idealities, demonstrating a practical and scalable path toward ultra-low-power NeSy hardware.
\end{abstract}
\keywords{Neuro-symbolic computing, Compute-in-memory (CiM), FeFET, Ultra-low power, CiM Simulator, Optimization.}

\maketitle

\section{Introduction}
\label{sec:introduction}

\begin{figure*}
    \centering
    \vspace{-0.5em}
    \includegraphics[width=1\linewidth]{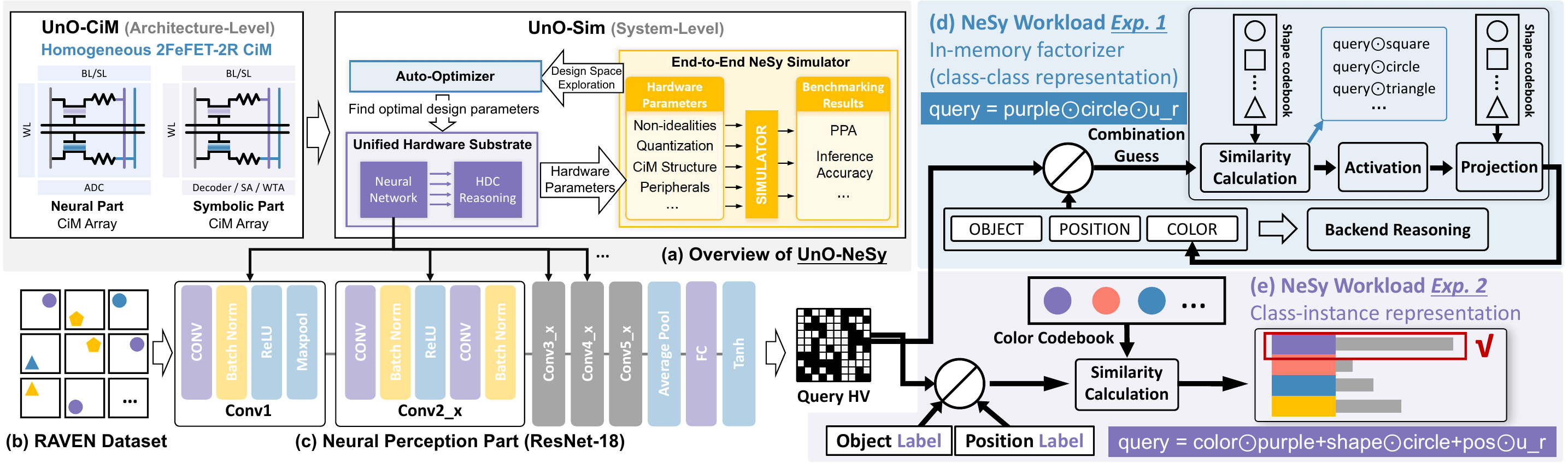}
    \vspace{-2.5em}
    \caption{(a) Overview of UnO-NeSy. 
    (b-e) Illustration of NeSy workloads.
    (b) The RAVEN dataset fed into the (c) ResNet-18 neural network for feature extraction.
    (d) \textbf{Exp.~1:} based on the original in-memory factorizer using class–class representations, where iterative similarity calculations and projections are performed to factorize the query HV.
    (e) \textbf{Exp.~2:} based on class–instance type of representations, which simplifies the factorization process by directly performing similarity calculation with label HVs.}
    \label{fig:overview}
    \vspace{-1em}
\end{figure*}

The safety-critical nature of autonomous driving requires AI systems to clearly and logically understand complex, dynamic environments \cite{kim2018textual, Yuan2024RAGDriver}.
Traditional neural networks excel at perceptual tasks such as object detection and scene segmentation, but they lack explicit logical reasoning and contextual understanding, both of which are essential for reliable decision-making.
Neuro-symbolic artificial intelligence (NeSy AI) combines neural perception with symbolic reasoning, providing a promising approach for interpretable and robust decision-making in safety-critical applications \cite{bhuyan2024neuro,wan2024towards,Langenegger2023,camposampiero2025raven,confalonieri2025multiple}. 
Hyperdimensional computing (HDC), inspired by the information processing patterns in human brains, offers a robust and noise-tolerant framework for symbolic reasoning and associative memory \cite{Plate1994, verges2025classification, zou2025hyperdimensional}.
HDC encodes information as hypervectors (HVs), enabling efficient symbolic operations such as binding and permutation.
These characteristics make HDC promising for implementing explainable and fault-tolerant NeSy AI \cite{zou2025hyperdimensional}.
However, both neural perception and HDC reasoning require intensive memory access and computation, resulting in substantial data-movement overhead on conventional von Neumann platforms.

Compute-in-memory (CiM) offers a promising solution by performing computation directly within memory arrays, significantly improving energy efficiency for data-intensive workloads such as neural networks and HDC \cite{qian2025ferroelectric, yin2024ferroelectric, li2025high, gtrcim, yan2023improving}.
Among various non-volatile memory (NVM) used in CiM, ferroelectric FETs (FeFETs) provide a unique opportunity for unified NeSy acceleration.
As voltage-controlled devices with high I$_{ON}$/I$_{OFF}$ ratio and CMOS compatibility \cite{hu2021memory,qian2024hycim,chen2025low, chakraborty2025ferroelectric, yin2025}, FeFETs enable flexible analog computation.

Despite these opportunities, prior NeSy works \cite{wan2024towards, Langenegger2023, du2025crosslayerdesignvectorsymboliccomputing} mainly focus on algorithm or architectural optimizations, with hardware implementations that are either digital or heterogeneous, thus failing to fully exploit CiM parallelism and FeFET device characteristics.
Moreover, the neural and symbolic stages in NeSy systems exhibit different hardware requirements. Neural perception is dominated by dense MAC operations and is more sensitive to hardware noise, whereas HDC reasoning involves binding and similarity calculation with distinct dataflows and different tolerance to device variations. Therefore, optimizing only a single operator or using a fixed hardware setting for the entire pipeline is insufficient.
Existing CiM simulators, such as NeuroSim \cite{read2025neurosim} and MNSIM \cite{mnsim}, mainly support neural workloads and cannot model the interaction between hardware non-idealities and NeSy operations. This lack of a unified hardware and evaluation framework limits the exploration of efficient FeFET-based NeSy CiM designs.

To address these challenges, we propose \textbf{UnO-NeSy}, a unified and hardware-aware CiM framework for optimizing NeSy computing. 
It consists of \textbf{UnO-CiM}, an ultra-low-power NeSy architecture based on a homogeneous 2FeFET--2R CiM array that unifies neural computation and HDC symbolic reasoning, and \textbf{UnO-Sim}, a cross-workload, end-to-end CiM simulator that enables accurate evaluation and automated design optimization. 
Fig.~\ref{fig:overview}(a) summarizes the key contributions:

\vspace{-0.5em}
\begin{itemize}

\item
At the architecture level, we propose \textbf{UnO-CiM}, an ultra-low-power NeSy architecture based on a homogeneous 2FeFET--2R CiM array. 
The array supports neural MAC operations, HDC binding, and similarity calculation within a unified hardware substrate.  
Leveraging the voltage-dependent behavior of FeFETs, the proposed design enables an adaptive noise--energy trade-off for NeSy computations.

\item
At the system level, we develop \textbf{UnO-Sim}, which is, to our knowledge, one of the first \textbf{cross-workload and end-to-end CiM simulators for NeSy} that jointly models neural quantization, device and circuit non-idealities, HDC execution behavior, and system-level hardware mapping. It enables consistent evaluation of inference accuracy, latency, energy, and area across neural perception and symbolic reasoning.

\item
Building on UnO-Sim, we develop an optimization framework for \textbf{UnO-NeSy}, enabling \textbf{automatic operating point optimization} for NeSy CiM systems under coupled energy, accuracy, and hardware constraints. 
Across representative workloads~\cite{Langenegger2023, kanerva2010}, the optimized design improves energy efficiency by $61.8\%$ while maintaining high inference accuracy compared with the baseline 2FeFET--2R design.

\end{itemize}
\vspace{-10pt}
\section{Background and Motivation}
\label{sec:background}


\subsection{NeSy Workloads and Hardware Challenges}

NeSy AI integrates human-comprehensible symbolic reasoning with neural perception to enable logical, interpretable decision-making \cite{wan2024towards,Sarker2022, zhou2025factorhd}.
Fig. \ref{fig:overview} presents the overall schematic of the HDC-based NeSy workloads considered in this work.
In this study, we focus on factorization, one of the fundamental reasoning tasks in NeSy AI.
As shown in Fig. \ref{fig:overview}(b-c), the RAVEN dataset \cite{Zhang_2019_CVPR} is first processed by a ResNet-18 neural network to extract visual features and generate  corresponding query HV for inference.
Fig. \ref{fig:overview}(d-e) illustrates two representative types of NeSy workloads using two primary HDC encodings:
\textbf{Exp. 1:} the class–class representation, where relations are encoded through binding–binding operations \cite{Hersche2023}.
Fig. \ref{fig:overview}(d) illustrates the in-memory factorizer \cite{Langenegger2023} that iteratively explores the probability space through similarity calculation and projection until finally finding out the correct attributes combination.
\textbf{Exp. 2:} the class–instance representation, where objects are encoded using binding–bundling operations \cite{kanerva2010}.
Fig. \ref{fig:overview}(e) shows a factorizer for class–instance relations, where query HVs can be directly factorized by binding with label HVs followed by similarity calculation, eliminating the need for thousands of iterations.

However, the neural perception component of NeSy AI requires extensive MAC operations across deep neural layers, while the symbolic reasoning module involves massive similarity calculations among HVs for associative reasoning \cite{fahdcim}. 
These two computational patterns lead to a "memory wall" problem in conventional von Neumann architectures caused by frequent and energy-intensive data transfers \cite{Wulf_1995}, highlighting the need for energy-efficient hardware solutions for NeSy AI systems.

\begin{figure}
    \centering
    \includegraphics[width=1\linewidth]{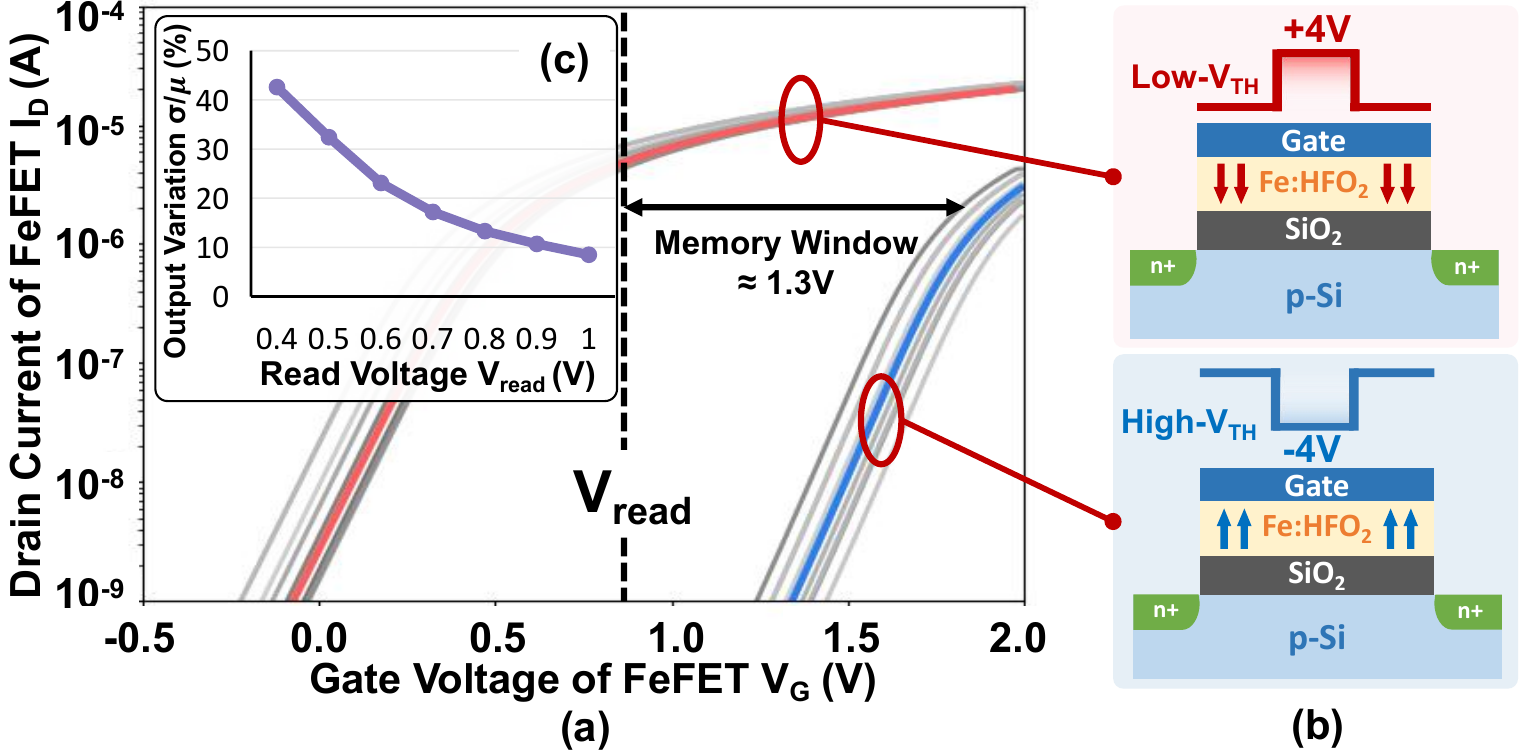}
    \vspace{-2.5em}
    \caption{
    (a) Characteristic  of FeFET with binary states in series with a resistor. The process variation affects $V_{\mathrm{TH}}$.
    (b) FeFET structure with binary states.
    (c) Output variation of 1FeFET-1R cell under different read voltages.
    }
    \label{fig:fefet}
    \vspace{-2em}
\end{figure}

\begin{figure*}
    \centering
    \includegraphics[width=1\linewidth]{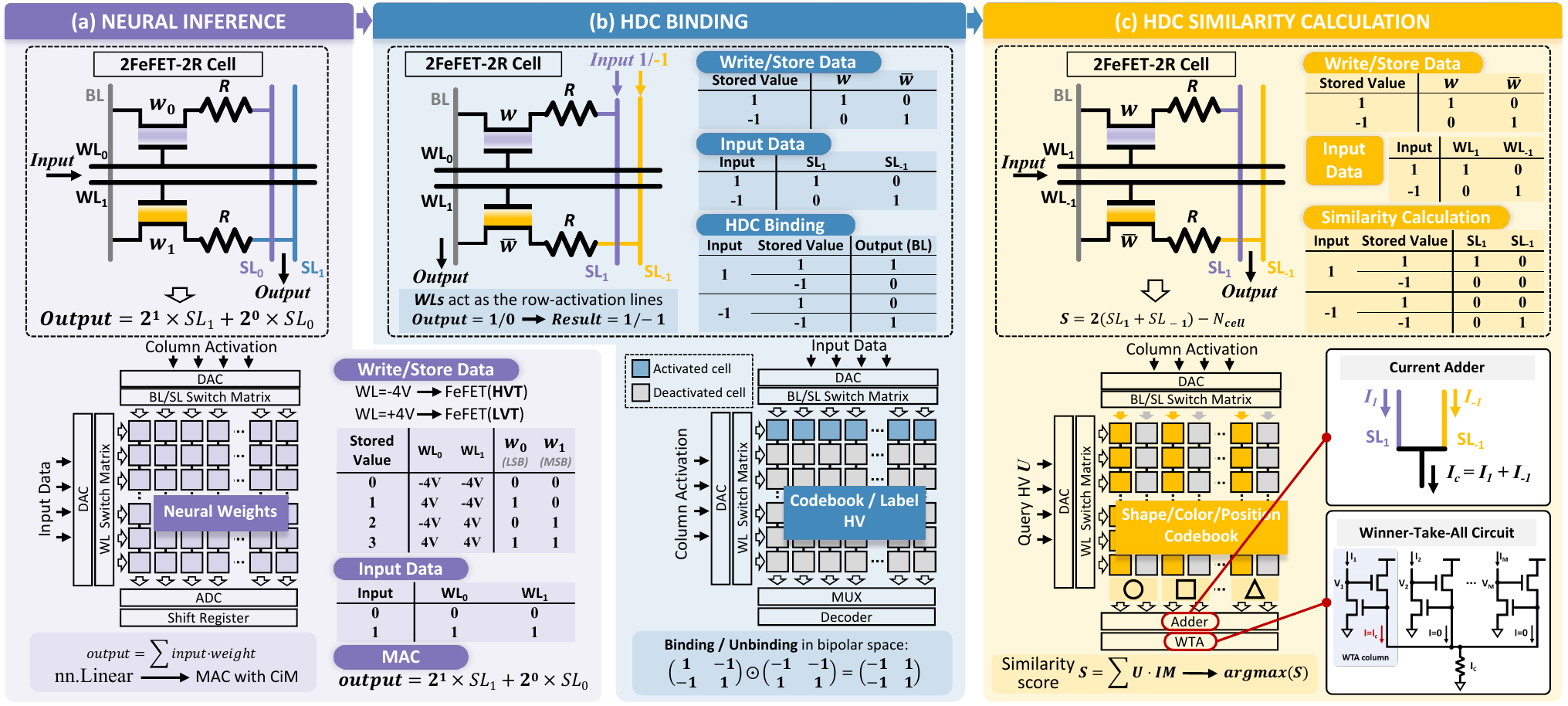}
    \vspace{-2.5em}
    \caption{ 
     (a) Neural network inference mapped to in-memory MAC operations.
     (b) Query HV unbinding through iterative binding with stored label or item HVs.
     (c) Factorization via dot-product similarity computation against attribute codebooks and winner-take-all (WTA) selection.
     }
    \label{fig:newarray}
    \vspace{-1.5em}
\end{figure*}

\subsection{CiM and FeFET Basics}
\label{sec:device}
To address the "memory wall" bottleneck, CiM employs parallel computation directly within memory arrays, emerging as a promising solution for energy-efficient designs \cite{qian2025ferroelectric, yin2024ferroelectric, li2025high, gtrcim, yin2024remedy, huang2023fefet}.
Its high parallelism also makes CiM particularly effective for HDC operations. 

FeFETs provide additional opportunities for NeSy acceleration due to their ultra-low leakage    current, high I$_{ON}$/I$_{OFF}$ ratio, voltage-driven mechanisms, and compatibility with CMOS technology \cite{qian2024hycim,ni2019ferroelectric, Yang2026, yin2024deepforest, qian2025device}. 
Fig. \ref{fig:fefet}(b) shows that, by applying positive or negative gate pulses ($\pm$4V) to FeFET devices, ferroelectric polarization switches to point towards the channel or the gate, setting the device to low-V$_{TH}$/high-V$_{TH}$ (LVT/HVT) state \cite{khan2020future}. The stored data ("1" or "0") can be read by measuring the drain current (i.e., I$_{ON}$ or I$_{OFF}$) while applying a lower gate voltage between the LVT and HVT, as shown in Fig. \ref{fig:fefet}(a).
Since HDC involves intensive read/write access, FeFETs are particularly suitable for low-power HDC implementations.

Scaling the operating voltage $V_{\mathrm{read}}$ into the subthreshold region provides an effective approach to improve the energy efficiency. 
Nevertheless, Fig.~\ref{fig:fefet}(c) illustrates that the output variation increases exponentially as $V_{\mathrm{read}}$ decreases, considering FeFET V$_{TH}$ variability of $\sigma_{V_{TH}}=54\,\mathrm{mV}$~\cite{Soliman_2020, hu2021memory} together with resistor variations. 
Therefore, adjusting $V_{\mathrm{read}}$ enables a tunable trade-off between energy consumption and output noise at the cell level.



\vspace{-5pt}
\section{Ultra-Low-Power UnO-NeSy Design}
\label{sec:schematic}


\subsection{UnO-CiM: Homogeneous 2FeFET-2R CiM}
\label{sec:proposal}


Existing CiM designs are primarily optimized for dense neural workloads and do not consider the mixed operation patterns in NeSy, such as symbolic HV factorization and binding operations. 
To support the distinct dataflows of neural perception and HDC reasoning,  we propose a homogeneous 2FeFET–2R CiM array design. The proposed design simultaneously supports neural MAC operations, HDC binding, and factorization, enabling end-to-end NeSy execution within a single CiM substrate.

The proposed 2FeFET–2R CiM cell is shown in Fig.~\ref{fig:newarray}. Each cell consists of two FeFETs, and each FeFET connects in series with a resistor $R$ to reduce output variation.
The two FeFETs share the same bitline (BL). The FeFET gates are driven by two word lines (WLs), and the sources are connected to different source lines (SLs). 
Owing to the identical 1FeFET-1R structure on each branch, the proposed cell exhibits equivalent output characteristics to a 1FeFET-1R cell on each SL.
The FeFETs are programmed to the LVT and HVT states using positive and 
negative write pulses, respectively.
The final cell state is determined by the joint state of the two FeFETs.
Detailed write schemes are shown in Fig.~\ref{fig:newarray}(a–c) for different operation modes. 
For example, in neural networks, the two FeFETs store $w_0$ and $w_1$ to represent 2-bit weights. In HDC operations, different state combinations ($w$, $\bar{w}$) are used to represent '-1' and '1'.

Fig.~\ref{fig:newarray}(a) illustrates the neural computing mode. 
Since each cell contains two FeFETs, up to 2-bit weights can be stored to fully utilize the structure. 
Input data are applied through WLs, where a positive $V_{read}$ represents '1' and 0 V represents '0'. BLs of activated columns are biased at 1V. Column currents are accumulated for MAC operations and digitized by ADCs.

Fig.~\ref{fig:newarray}(b) illustrates the HDC binding operation of bipolar HVs. 
One bipolar HV is stored using complementary FeFET states, while the other HV is applied through $SL$s. 
During operation, rows are activated via setting both WLs of each row to 1V. 
For inputs, values '1' and '-1' are applied through different voltage levels on $SL_1$ and $SL_{-1}$. For example, input '1' is applied by setting $SL_1$ to 1V, while input '-1' is applied through $SL_{-1}$.
Matching stored and input values generate current on BL, producing the 
element-wise binding output.
Each mapped row is sequentially activated to generate element-wise binding currents, which are sensed and decoded to reconstruct the complete HV binding result.
Binding adopts a row-wise execution scheme, which is sufficient for HDC workloads where binding operations are typically performed sequentially.

\begin{figure*}
    \centering
    \includegraphics[width=1\linewidth]{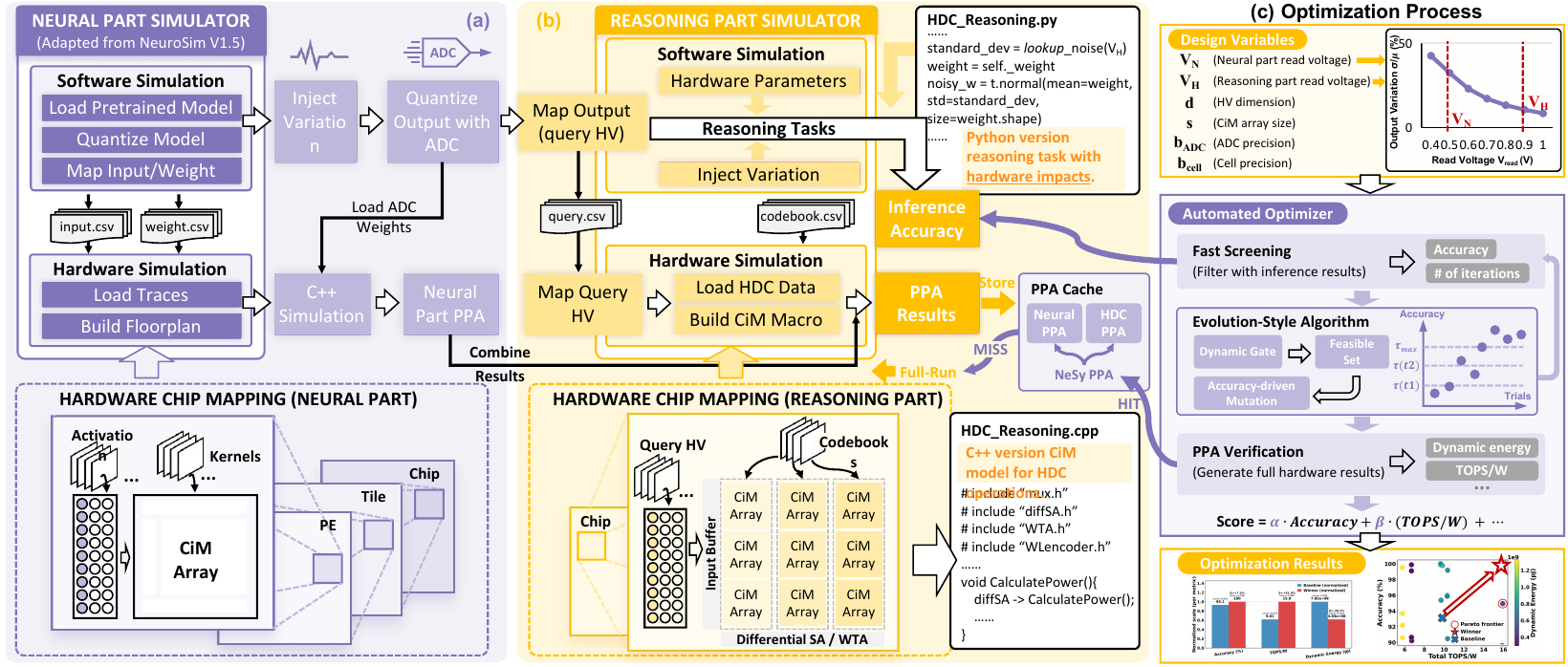}
    \vspace{-2.5em}
    \caption{
    (a) The proposed UnO-Sim simulator applies revised NeuroSim for neural part simulation. 
    (b) For HDC reasoning part, UnO-Sim maps codebooks to multiple CiM arrays. The query HV generated by the neural part is  sent to CiM arrays as the input to perform binding and similarity computation. 
    (c) Optimization framework for NeSy CiM. The optimizer first performs fast accuracy-driven screening with an annealed constraint, then conducts detailed PPA evaluation on selected candidates.
    }
    \label{fig:simulator}
    \vspace{-1em}
\end{figure*}

Fig.~\ref{fig:newarray}(c) illustrates HDC similarity calculation based on dot products. Code vectors are stored column-wise and can be partitioned across multiple columns, with their partial similarities accumulated before WTA selection. 
During inference, the query HV is applied to the WLs. For example, input '1' is applied by setting $WL_1$ to a positive $V_{read}$, while input '-1' is applied by setting $WL_{-1}$ to $V_{read}$. 
In this work, V$_{read}$ is set as low as 0.4 V, and peripheral circuits are also adapted to support subthreshold computing.
BL signals activate the selected columns. 
Similarity calculations for all items within each codebook are performed in parallel.
The resulting SL currents represent the similarity between the query HV and stored vectors.
The similarity score is
$S=2(\sum I_{\mathrm{SL}_1}+\sum I_{\mathrm{SL}_{-1}})-N_{\mathrm{cell}}$, where $N_{\mathrm{cell}}$ is the constant number of cells per column and can therefore be omitted.
A winner-take-all (WTA) circuit \cite{lazzaro1988winner} then identifies the maximum-similarity column without requiring high-resolution ADCs, significantly reducing energy consumption. 

\vspace{-5pt}
\subsection{UnO-Sim: NeSy CiM Hardware Simulator}
\label{subsec:nesy_sim}

Designing NeSy accelerators with analog CiM arrays requires jointly modeling neural feature extraction and symbolic reasoning under device/circuit non-idealities, limited peripheral precision, and system-level mapping constraints. 
However, SPICE-level simulation of end-to-end NeSy pipelines is intractable, while existing CiM simulators primarily target DNN inference and lack support for reasoning primitives. 
To address this gap, we develop UnO-Sim, an end-to-end NeSy CiM simulator that enables unified evaluation of inference accuracy and PPA across the full NeSy pipeline. 
The simulator can be adapted to various NeSy architectures.

At the software level, UnO-Sim performs inference while incorporating hardware effects, including quantization and device/circuit variations, which are consistently modeled across the entire pipeline. 
Neural outputs are directly propagated to the reasoning stage. 
The inference process produces the prediction accuracy and the number of reasoning iterations \(N_{\mathrm{iter}}\).
At the hardware level, neural and reasoning workloads are mapped onto separate CiM arrays. 
As shown in Fig.~\ref{fig:simulator}(a) and (b), the neural module is revised from NeuroSim~\cite{read2025neurosim}, while the reasoning module incorporates dedicated peripheral configurations for HDC operations. 
The proposed 2FeFET--2R cell and HDC-specific peripheral circuits are characterized using SPICE simulations and integrated into UnO-Sim.
UnO-Sim independently models the read voltages of the neural and reasoning stages using Virtuoso-calibrated noise and PPA characteristics.
During accuracy evaluation, each candidate’s \(V_N\) or \(V_H\) is mapped to an neural or HDC noise standard deviation using the voltage-noise lookup table.
Due to the strong coupling between operating voltage, noise, and energy efficiency in FeFET-based CiM, this decoupling enables exploration of heterogeneous operating regimes. 
For example, neural inference operates in near-saturation regions for improved stability, while reasoning operates in the subthreshold regime for ultra-low power consumption, fully exploiting the voltage-tunable characteristics of FeFET devices. 
During PPA evaluation, different read voltages are applied, and reasoning cost is scaled by \(N_{\mathrm{iter}}\). 
The overall PPA is obtained by aggregating the energy, delay, and area contributions of both neural and reasoning modules.
Since the neural and reasoning hardware costs are separable, UnO-Sim caches the evaluated PPA of each module and recombines them across design candidates, invoking detailed hardware simulation only for unseen configurations.



\vspace{-5pt}
\subsection{Automatic Hardware Design Optimization for UnO-NeSy}
Building on UnO-Sim, we develop an automatic accuracy-constrained co-optimization 
framework to jointly explore the NeSy CiM configuration
\(\mathbf{x}=\{V_{\mathrm{N}}, V_{\mathrm{H}}, d, s, b_{\mathrm{ADC}}, b_{\mathrm{cell}}\}\), 
where $V_{\mathrm{N}}$ and $V_{\mathrm{H}}$ denote the read voltages of the 
neural and reasoning stages, respectively, $d$ is the HV dimension, $s$ is 
the CiM array size, and $b_{\mathrm{ADC}}$ and $b_{\mathrm{cell}}$ denote the 
ADC and cell precision. 
For iterative NeSy workloads, the task-level dynamic energy is 
modeled as $E_{\mathrm{total}}(\mathbf{x})=
E_{\mathrm{N}}(\mathbf{x})+
N_{\mathrm{iter}}(\mathbf{x})E_{\mathrm{H}}(\mathbf{x})$, where 
$E_{\mathrm{N}}$ is the neural-stage energy and $E_{\mathrm{H}}$ is the 
reasoning energy per iteration. Task-level latency is calculated in the 
same manner.
Candidates are ranked using
\vspace{-0.5em}
\begin{equation}
S(\mathbf{x})
=
\alpha \widetilde{A}(\mathbf{x})
+
\beta \widetilde{\mathrm{EE}}(\mathbf{x})
+
\gamma
\left[
1-\widetilde{E}_{\mathrm{total}}(\mathbf{x})
\right],
\quad
A(\mathbf{x}) \geq A_{\min},
\vspace{-0.5em}
\end{equation}
where $\widetilde{A}$, $\widetilde{\mathrm{EE}}$, and 
$\widetilde{E}_{\mathrm{total}}$ denote the normalized inference accuracy, energy efficiency, and dynamic energy, respectively. 

As illustrated in Fig.~\ref{fig:simulator}(c), the optimization proceeds in 
four steps. \textbf{1) Fast screening:} end-to-end inference is performed to estimate the accuracy and $N_{\mathrm{iter}}$ of each candidate without invoking full PPA simulation.
\textbf{2) Feasible-set construction:} a dynamically tightened accuracy gate selects feasible parent configurations; 
when the feasible set is empty, the highest-accuracy candidates are retained as surrogate parents. 
\textbf{3) Hardware-aware mutation:} when accuracy is insufficient, mutations favor hardware configurations with higher representational capacity or improved signal quality; 
after satisfying the accuracy gate, the search shifts toward lower-energy configurations. 
Candidates satisfying the accuracy requirement are sent for further PPA validation.
\textbf{4) PPA verification:} selected candidates undergo high-fidelity PPA evaluation, reusing cached module results when available. The candidate with the highest score $S(\mathbf{x})$ is selected as the final configuration.
\vspace{-5pt}
\section{Performance Evaluations}

\subsection{Experimental setup}
We calibrate UnO-Sim using the analyzed SPICE-derived 2FeFET--2R cell characteristics.
The automatic optimization framework is evaluated on \textbf{Exp.~1} and \textbf{Exp.~2}. 
For the neural backbone, we adopt ResNet-18 following the standard ImageNet configuration for feature extraction.
In \textbf{Exp.~1}, the stride of the first convolution layer is modified to 1.
Both neural inputs and weights are quantized to 8-bit precision. 
For inference accuracy evaluation, we consider an experimental FeFET threshold voltage variability of $\sigma_{V_T}=54\,\mathrm{mV}$~\cite{Soliman_2020} and resistors of $R=30\,\mathrm{k}\Omega$ with $10\%$ variation. 
We perform 500 Monte Carlo simulations across different operating voltages, resulting in normalized output deviations (e.g., $\sigma/\mu=10.7\%$ at $V_{\mathrm{read}}=0.9\,\mathrm{V}$) for the proposed 2FeFET–2R cell. 
For hardware modeling, parameters of the 45 nm 2FeFET--2R cell are extracted from SPICE simulations, including an on/off ratio of 230 and voltage-dependent noise characteristics across operating regimes.
We evaluate inference accuracy and hardware performance on the RAVEN dataset~\cite{Zhang_2019_CVPR}, which consists of objects characterized by attributes such as type, size, color, and position. 
The first codebook represents the position attribute, the second represents color, and the third combines size and type attributes, resulting in 30 size-type combinations.
We first specifically evaluate on the center-single pattern of RAVEN.
The explored design space is 
$\mathbf{x}=\{V_{\mathrm{N}}, V_{\mathrm{H}}, d, s, b_{\mathrm{ADC}}, b_{\mathrm{cell}}\}$,
where $V_{\mathrm{N}}\in\{0.7,0.8,0.9,1.0\}$, 
$V_{\mathrm{H}}\in\{0.4,0.5,\dots,1.0\}$, 
$d\in\{128,256,512\}$, 
$s\in\{128,256\}$, 
$b_{\mathrm{ADC}}\in\{6,7,8,9\}$, 
and $b_{\mathrm{cell}}\in\{1,2\}$.
The baseline 2FeFET-2R CiM design is set with $\{V_{\mathrm{N}}=1V, V_{\mathrm{H}}=1V, d=256, s=128, b_{\mathrm{ADC}}=7, b_{\mathrm{cell}}=1\}$ for both NeSy workloads.
The optimization objective is to achieve high inference accuracy with improved energy efficiency under the above design constraints.
For \textbf{Exp.~1}, the scoring weights are set to
$(\alpha,\beta,\gamma)=(0.5,0.3,0.2)$ to account for the additional impact 
of iterative reasoning on task-level energy. For \textbf{Exp.~2}, where 
$N_{\mathrm{iter}}=1$ for all candidates and thus energy-related metrics exhibit similar trends, we use
$(\alpha,\beta,\gamma)=(0.5,0.5,0)$ to prioritize inference accuracy and 
energy efficiency.

We then benchmark the inference accuracy of UnO-NeSy on all patterns of the RAVEN dataset, focusing on the factorization task under \textbf{Exp.~2}.
We additionally include the NVSA backend \cite{Hersche2023} to evaluate the full RAVEN reasoning workload. 
Because both the \textbf{Exp.~1} and \textbf{Exp.~2} workloads support only single-object factorization, input images are partitioned during preprocessing.
The inference accuracy is compared against representative NeSy designs such as software-level NVSA \cite{Hersche2023}, a conventional digital NeSy implementation \cite{wan2024towards} and the algorithm-level PrAE model \cite{zhang2021abstract}.

We further benchmark UnO-NeSy on more complex factorization tasks, following the setup of the in-memory factorizer~\cite{Langenegger2023}, i.e., $D=512$ and $M=512$. 
In this setting, UnO-NeSy is evaluated under both \textbf{Exp.~1} and \textbf{Exp.~2}. 
The baseline 2FeFET--2R CiM design is configured as 
$\{V_{\mathrm{N}}=1\,\mathrm{V}, V_{\mathrm{H}}=1\,\mathrm{V}, d=512, s=256, b_{\mathrm{ADC}}=8, b_{\mathrm{cell}}=1\}$. 
We further apply the proposed optimizer to identify improved operating points for the 2FeFET--2R CiM design.  
We compare the performance of the proposed design in both HDC similarity computation and end-to-end NeSy inference against prior in-memory factorization approaches, including a PCM-based design~\cite{Langenegger2023} and a RRAM-based CiM design~\cite{Wantong2022}.

\begin{figure*}
    \centering
    \includegraphics[width=1\linewidth]{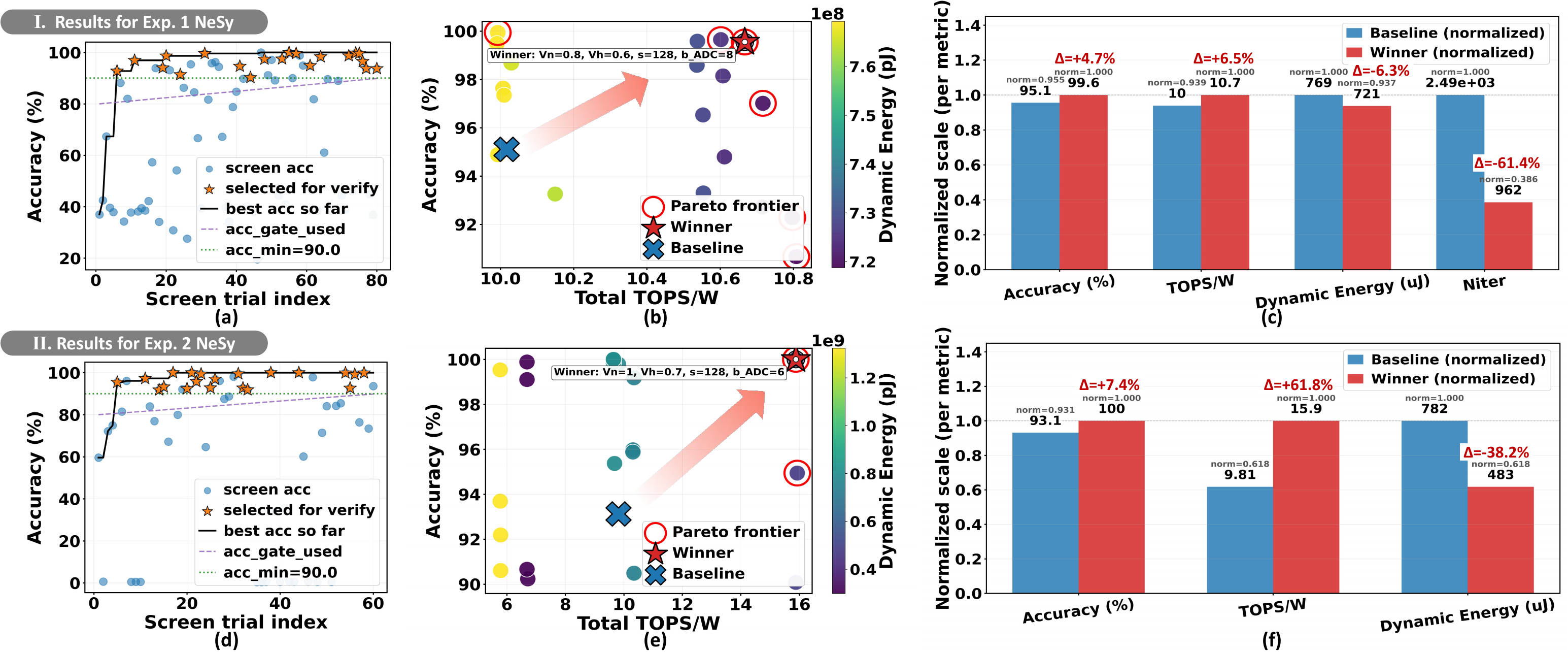}
    \vspace{-2.5em}
    \caption{
    Optimization results for (a--c) \textbf{Exp. 1} and (d--f) \textbf{Exp. 2} NeSy workloads using the proposed automatic optimizer on the proposed 2FeFET-2R CiM design. 
    (a, d) Optimization timeline: screening-stage accuracy versus trial index, with selected candidates marked for verification. 
    (b, e) Optimization map showing all the verified points: accuracy versus total TOPS/W, with color indicating dynamic energy. 
    (c, f) Winner–baseline comparison.
    For \textbf{Exp. 2}, $N_{\mathrm{iter}}=1$ and is therefore omitted.
    }
    \label{fig:optimized}
    \vspace{-1em}
\end{figure*}

\vspace{-5pt}
\subsection{Automatic energy--accuracy optimization}

\begin{table}[t!]
\centering
\caption{Performance of 2FeFET-2R cell on different operators}
 \vspace{-3mm}
\label{tab:operator}
\centering
    \includegraphics[width=1\linewidth]{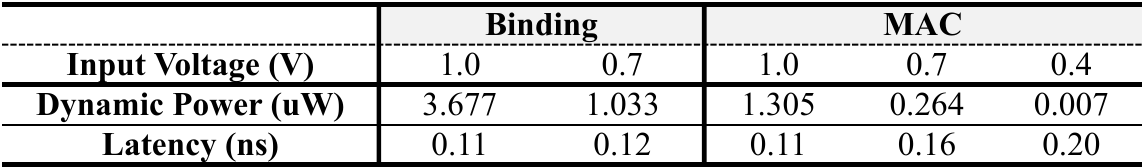}
    \vspace{-2.5em}
\end{table} 

We first evaluate the performance of the proposed 2FeFET-2R cell design. 
Table~\ref{tab:operator} shows that voltage scaling substantially reduces cell-level power with limited latency overhead.
Peripheral circuits are implemented in 45nm technology, and simulations show that each WTA column consumes 60\,\textmu W.
At the macro level, the CiM array, ADC and precharge circuits, accumulation circuits, and other peripherals account for $32.3\%$, $32.0\%$, $8.5\%$, and $27.2\%$ of the total area, respectively, while the WTA occupies less than $10\%$ of the ADC area.
Based on these evaluation results, we calibrate UnO-Sim for system-level simulation.

We then evaluate the effectiveness of the proposed optimizer in exploring the energy--accuracy trade-off for NeSy CiM systems. 

For the \textbf{Exp. 1} NeSy workload, 
Fig.~\ref{fig:optimized}(a) illustrates the optimization trajectory, where the screen-stage accuracy is plotted over trials. 
The selected best configuration (winner) is highlighted in Fig.~\ref{fig:optimized}(b), with the operating point 
$\{V_{\mathrm{N}}=0.8\,\mathrm{V}, V_{\mathrm{H}}=0.6\,\mathrm{V}, d=256, s=128, b_{\mathrm{ADC}}=8, b_{\mathrm{cell}}=2\}$.
Alternative scoring weights select the same or neighboring Pareto-optimal configurations.
The optimizer consumes less than $3\%$ of the runtime of brute-force search.
As summarized in Fig.~\ref{fig:optimized}(c), the optimized design improves accuracy from $95.1\%$ to $99.6\%$, increases energy efficiency by $6.5\%$, and reduces the total dynamic energy by $6.3\%$. 
In addition, the number of reasoning iterations is reduced by $61.4\%$, significantly lowering the total inference time and energy costs. 
By incorporating the measured number of reasoning iterations \(N_{\mathrm{iter}}\), the framework favors configurations that reduce task-level computation rather than optimizing hardware metrics in isolation.

To validate generality, we further evaluate the proposed method on the \textbf{Exp. 2} workload. 
Fig.~\ref{fig:optimized}(d--f) present the corresponding optimization results. 
As shown in Fig.~\ref{fig:optimized}(e), the optimizer again identifies improved configurations over the baseline.
The winner achieves an accuracy improvement from $93.1\%$ to $100\%$, along with a $61.8\%$ increase in energy efficiency and a $38.2\%$ reduction in dynamic energy, as summarized in Fig.~\ref{fig:optimized}(f).
Overall, the proposed optimization framework efficiently identifies high-quality configurations across different NeSy settings, demonstrating its effectiveness for automatic hardware co-design.

\begin{table}
    \centering
    \caption{Benchmarking results on RAVEN reasoning tasks}
    \vspace{-3mm}
    \includegraphics[width=1\linewidth]{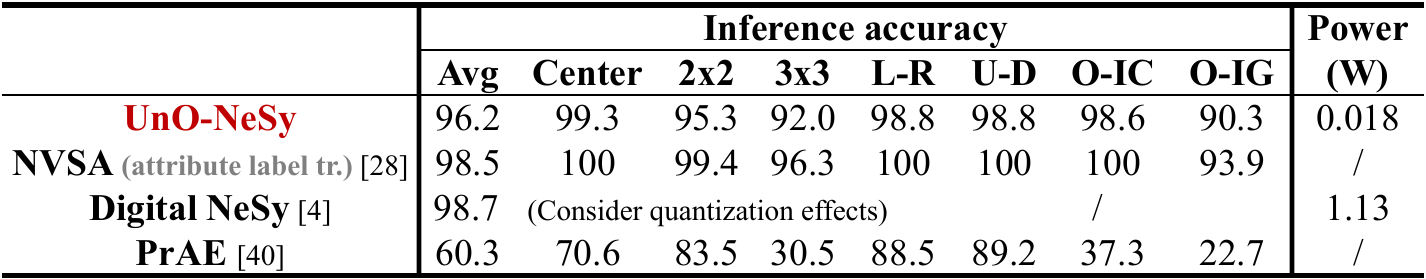}
    \label{tab:benchraven}
    \vspace{-3em}
\end{table}

We then use the winner configuration for \textbf{Exp.~2} NeSy to perform further reasoning tasks based on the factorization results. 
Table~\ref{tab:benchraven} summarizes the benchmarking results across multiple RAVEN patterns. 
Under modeled hardware non-idealities, UnO-NeSy incurs an average accuracy drop of only $2.3$ percentage points relative to software NVSA \cite{Hersche2023}.
Compared with the digital NeSy implementation \cite{wan2024towards}, it reduces power consumption by $98.4\%$ with only a $2.5$-percentage-point accuracy reduction, demonstrating its ultra-low-power advantage. 
Across all dataset types, UnO-NeSy also significantly outperforms PrAE~\cite{zhang2021abstract}. 
These results indicate that UnO-NeSy preserves high reasoning accuracy under realistic hardware conditions while significantly improving energy efficiency.



\vspace{-5pt}
\subsection{Benchmarking results}

\begin{table}
    \centering
    \caption{Benchmarking of UnO-NeSy on NeSy workloads}
    \vspace{-3ex}
    \includegraphics[width=1\linewidth]{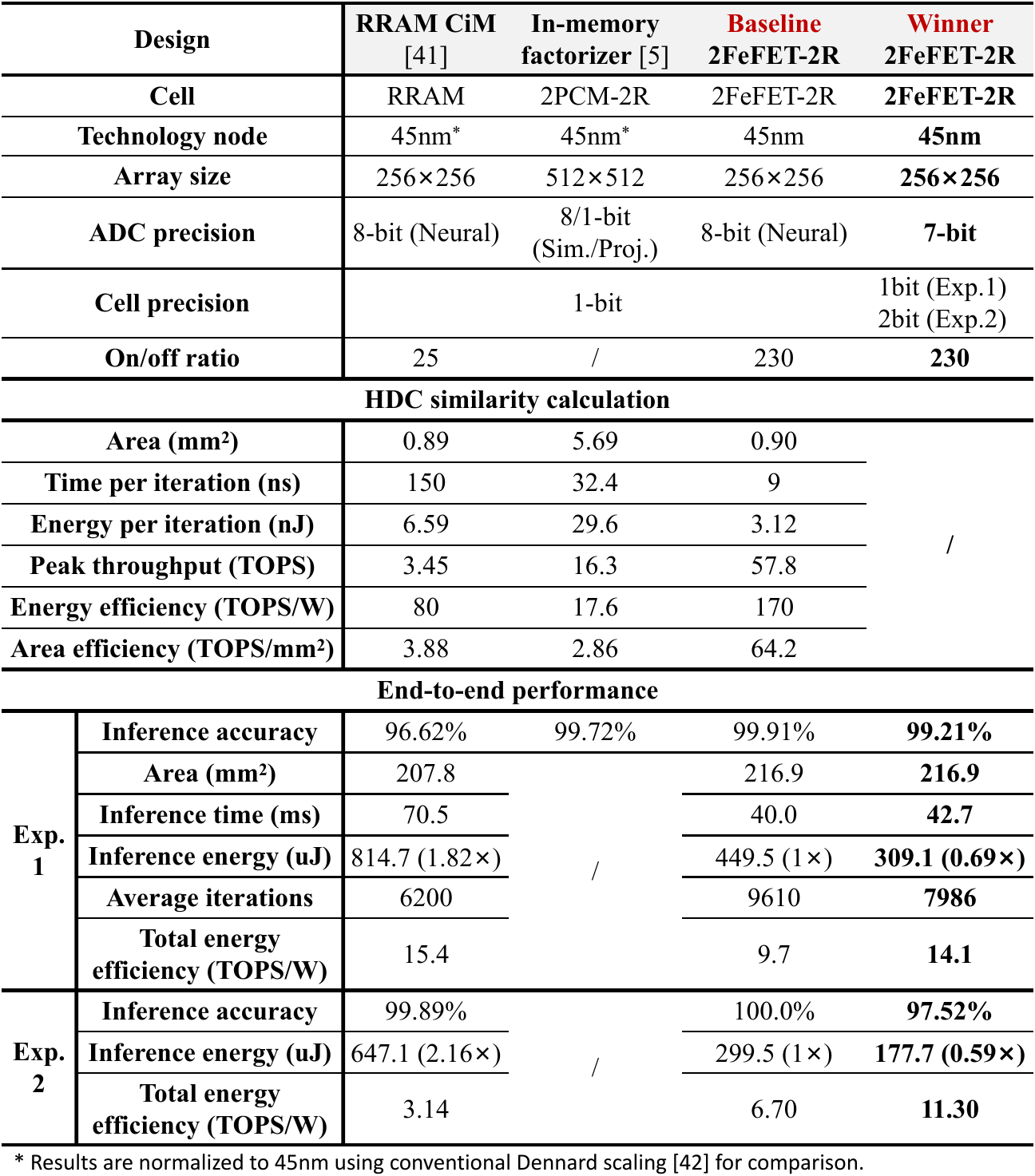}
    \label{tab:benchmark}
    \vspace{-2em}
\end{table}

Table~\ref{tab:benchmark} compares the baseline and optimizer-selected 2FeFET--2R configurations under the large-scale factorization setting for both the \textbf{Exp. 1} and \textbf{Exp. 2} workloads. \nocite{dennard1974design}
For the similarity calculation part that only involves HDC, the proposed baseline 2FeFET-2R CiM achieves $9.6\times$ the energy efficiency of the 2PCM--2R in-memory factorizer and $2.1\times$ that of the RRAM CiM.
For \textbf{Exp.~1}, when considering both the neural and reasoning parts, the baseline 2FeFET-2R CiM achieves an overall 9.7 TOPS/W energy efficiency. 
Using the optimizer, the winner 2FeFET-2R CiM with $\{V_{\mathrm{N}}=1\,\mathrm{V}, V_{\mathrm{H}}=0.7\,\mathrm{V}, d=512, s=256, b_{\mathrm{ADC}}=7, b_{\mathrm{cell}}=1\}$ significantly increases energy efficiency to 14.1 TOPS/W.
The total inference energy is reduced to 0.69× that of the baseline 2FeFET–2R CiM and 0.38× that of the RRAM-based CiM design.
Meanwhile, UnO-NeSy maintains $99.21\%$ inference accuracy, only $0.51$ percentage points lower than the in-memory factorizer.
For \textbf{Exp.~2}, factorization requires only a single iteration, making overall performance dominated by the neural component. 
In this case, the optimized 2FeFET--2R CiM with $\{V_{\mathrm{N}}=0.7\,\mathrm{V}, V_{\mathrm{H}}=1\,\mathrm{V}, d=512, s=256, b_{\mathrm{ADC}}=7, b_{\mathrm{cell}}=2\}$ reduces inference energy to $0.59\times$ that of the baseline while maintaining $97.52\%$ inference accuracy. 
Although the accuracy is slightly lower than that of the baseline, the optimized design achieves the highest score under the given optimization objective. 
Therefore, the HDC-only comparison reflects the architectural benefit of the proposed 2FeFET--2R CiM, while the winner-to-baseline gains quantify the contribution of the optimizer.

\vspace{-5pt}
\section{Conclusion}
\label{sec:conclusion}
This work presents \textbf{UnO-NeSy}, a unified CiM framework for optimizing NeSy computing. 
UnO-NeSy integrates \textbf{UnO-CiM}, a homogeneous 2FeFET--2R-based CiM architecture that unifies neural and symbolic computation within the same hardware substrate, and \textbf{UnO-Sim}, an end-to-end CiM simulator that enables accurate system-level evaluation under device non-idealities and quantization effects. 
Built on UnO-Sim, the proposed optimizer automatically explores workload-specific hardware configurations under coupled accuracy and energy constraints.
Compared with the baseline 2FeFET--2R configuration, the optimized design improves energy efficiency by $61.8\%$ and reduces dynamic energy by $38.2\%$.
Across various RAVEN configurations, UnO-NeSy achieves $96.2\%$ average
reasoning accuracy under realistic hardware conditions. 
These results highlight the effectiveness of homogeneous FeFET CiM and cross-stage optimization for efficient NeSy inference.

\begin{acks}
\vspace{-1ex}
This work was  supported by Zhejiang Provincial Key R\&D Program (2025C01071).
\end{acks}

{
\bibliographystyle{IEEEtran}
\bibliography{bib}
}

\end{document}